\documentclass[prb,preprint,showkeys,showpacs]{revtex4}
\usepackage{graphicx}
\usepackage{amssymb}
\usepackage{bm}
\usepackage{amsmath}

\begin{document}
\title{Plasmonic electromagnetically-induced transparency in symmetric structures}
\author{Xingri Jin}
\author{Yuehui Lu}
\author{Haiyu Zheng}
\author{YoungPak Lee}
\email{yplee@hanyang.ac.kr}
\thanks{Corresponding author}
\affiliation{Quantum Photonic Science Research Center and Department
of Physics, Hanyang University, Seoul 133-791, Republic of Korea}
\author{Joo Yull Rhee}
\email{rheejy@skku.edu}
\thanks{Co-corresponding author}
\affiliation{Department of Physics, Sungkyunkwan University, Suwon
440-746,  Republic of Korea}
\author{Won Ho Jang}
\affiliation{Korea Communication Commission Radio Research
Laboratory, Seoul 140-848, Republic of Korea}

\begin{abstract}
A broken symmetry is generally believed to be a prerequisite of
plasmonic electromagnetically-induced transparency (EIT), since the
asymmetry renders the excitation of the otherwise forbidden dark
mode possible. Nevertheless, according to the picture of
magnetic-plasmon resonance (MPR) mediated plasmonic EIT, we show
that the plasmonic EIT can be achieved even in the symmetric
structures based on the second-order MPR. This sharpens our
understanding of the existing concept, but also a profound insight
into the plasmonic coherent interference in the near-field zone.
\end{abstract}


\maketitle

Electromagnetically-induced transparency (EIT) is an important
quantum interference effect induced by the interaction between the
laser beams and atom ensembles under two-photon resonance
condition.\cite{Boller1991PRL, Harris1997phystoday,
Fleischhauer2005PMP} This effect can be applied to quantum
communication,\cite{LukinPRL2000} slow-light, and enhanced nonlinear
effects.\cite{HarrisPRL1999} Compared to the EIT of atomic system,
the plasmonic EIT in metamaterials has the advantages, such as
manipulation at room temperature, excitation of single optical
field, integration of nanoplasmonic circuits. Therefore, a great
deal of attention has been paid to the classical analogue of EIT in
mechanical oscillators, \emph{RLC} circuits,\cite{AlzarAJP2002}
optical resonators,\cite{OpatrnyPRA2001} optical dipole
antennas,\cite{Zhang2008PRL,Xu2009,Lu2009,LiuNM2009} trapped-mode
patterns,\cite{PapasimakisPRL2008} split-ring
resonators,\cite{TassinPRL2009,SinghPRB2009} and array of metallic
nanoparticles.\cite{YannopapasPRB2009}

Thanks to this a merging of plasmonics and metamaterials, it is of
great perspective to manipulate light at the nanometer scale with
metal nanostructures as nano-optical
components.\cite{Zhang2008PRL,ProdanSci2003} The application of
optical dipole antennas is a specific example of this merging among
the aforementioned studies on the classical analogue of EIT. Zhang
\emph{et al.}\cite{Zhang2008PRL} first proposed a scheme consisting
of bright and dark plasmonic modes that resembles the atomic system
of three levels. Subsequently, it was developed as a tripod system
manifesting the classical analogue of quantum coherence
swapping.\cite{Xu2009} Recently, Liu \emph{et al.}\cite{LiuNM2009}
experimentally demonstrated the plasmonic EIT at the Drude damping
limit using a stacked optical metamaterial composed of an upper gold
strip and a lower pair of gold strips with a dielectric spacer. It
was found that the asymmetry is a prerequisite of plasmonic EIT;
otherwise, only a single absorption peak is visible without any sign
of an EIT-like effect. Most of researchers also hold this
view\cite{LiuNM2009,SinghPRB2009,HaoNanolett2008} explicitly or
implicitly, since the dark mode is unlikely to be excited if it does
not have the broken symmetry.

In this letter we propose a scheme for the generation of plasmonic
EIT even in symmetric structures. This scheme is a minor
modification of the symmetric structure in Ref. 11 that makes the
EIT-like effect available. The underlying origin is also elucidated
in detail based on the picture of magnetic-plasmon resonance (MPR)
mediated plasmonic EIT.

\begin{figure}[t!]
\includegraphics[width=12.0cm]{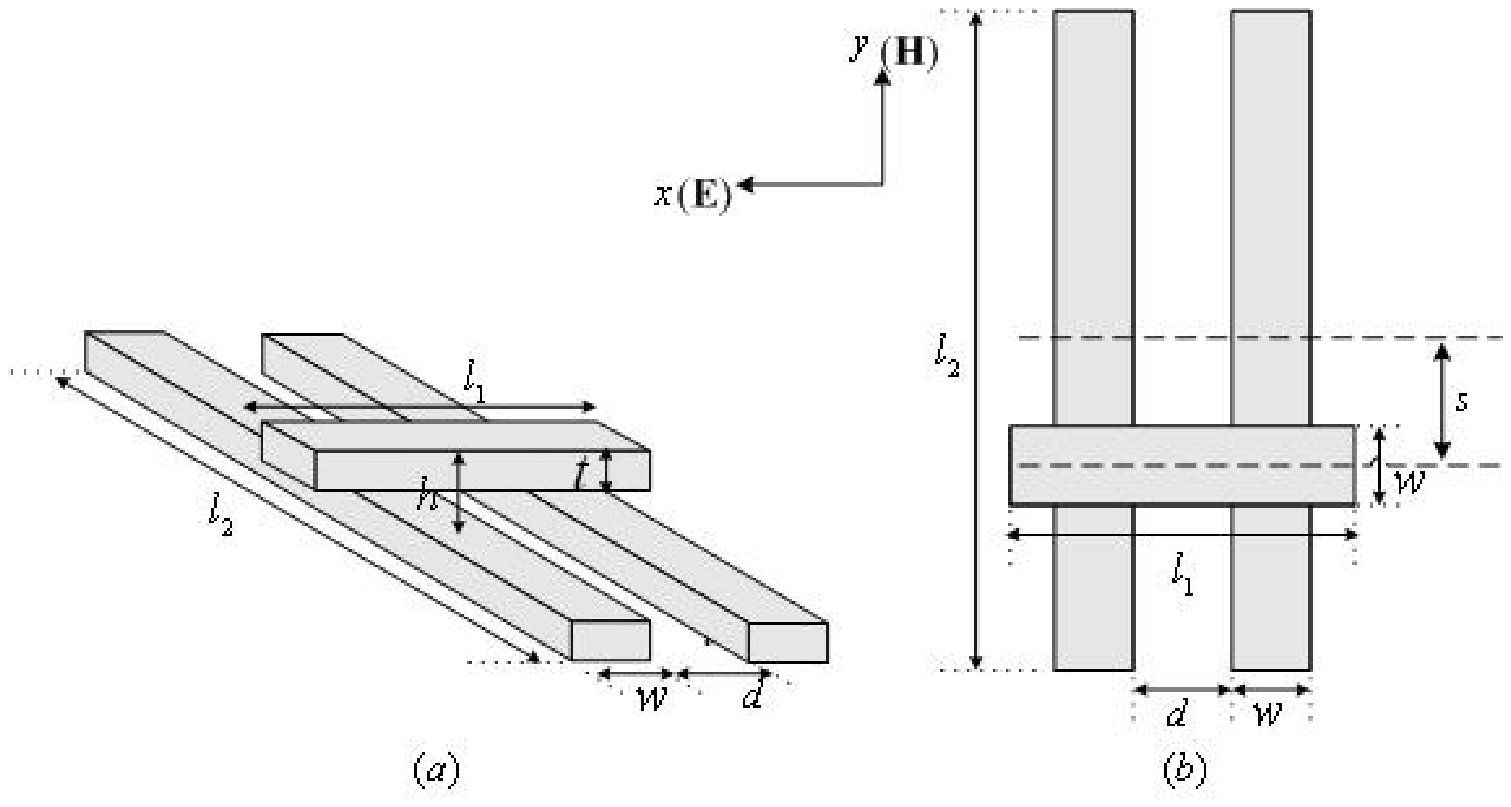}
\caption{Jin \emph{et al.}} \label{fig.1}
\end{figure}

As described in Ref. 11, each unit cell consists of an upper gold
strip as a bright mode and a lower pair of gold strips as a dark
mode with a dielectric spacer, as shown in Fig. \ref{fig.1}. In
particular, the parameter $s$ is defined to depict the lateral
displacement. Either the symmetry or the asymmetry is expressed as
$s = 0$ or $s \neq 0$, respectively. The plasmonic EIT originates
from the coupling of the two modes when the symmetry is
broken.\cite{LiuNM2009,SinghPRB2009,MaierNM2009} Essentially, the
former serves as an optical dipole antenna, and the latter as a
quadrupole antenna, when light is illuminated perpendicularly and
the electric field of the light is parallel to the upper strip. Our
study is limited to investigation of the symmetric structure, and
the similar geometrical parameters are used with two major
differences. One is that the dielectric spacer and the substrate are
not taken into account for simplicity, \emph{i.e.}, they are treated
as air, which does not affect the EIT-like feature except for a
blueshift. It does not cause any loss of generality in the
discussion afterwards. The other is that the lower pair of gold
strips is elongated to 790 nm, about two times longer than that in
Ref. 11. The reason for this elongation will be explained later. The
numerical calculation is carried out by using a finite-integration
package, CST Microwave Studio. The permittivity of gold is described
by the Drude model with a plasmon frequency of $\omega_p =
2\pi\times2.175\times10^{15}$ rad/s and a collision frequency of
$\nu_c = 1.225\times10^{14}$ Hz, which is three times larger than
that in bulk gold. \cite{LiuNM2009,DollingSci2006}

Figure \ref{fig.2} displays the simulated transmission and
absorption spectra for the symmetric structures with the different
vertical distance $h$. The latter is calculated according to $A = 1
- T - R$, where $T$ and $R$ denote the transmission and the
reflection, respectively. Astonishingly, the EIT-like feature
completely disappears when the length of the lower pair is
relatively short, $l_2 = 315$ nm, without varying the other
parameters. In contrast, this feature clearly manifests around 240
THz if the lower pair is elongated to 790 nm without any other
variations. In other words, whether the plasmonic EIT can be excited
or not depends on the length of the lower pair, rather than the
structural asymmetry. This appears inconsistent with the conclusion
in Ref. 11, where it is believed that the coupling fades away owing
to the structural symmetry. Therefore, to solve this puzzle, what
the coupling is and how the coupling works must be deciphered. Lu
\emph{et al.}\cite{Lu2009} provided a physical picture for plasmonic
EIT, in which it is considered as a result of plasmonic coherent
interference in near-field zone based on the excitation of surface
plasmon polaritons (SPPs) and MPR. The former is formed on the upper
strip since it behaves as an optical dipole
antenna,\cite{NovotnyPRL2007} while the latter is induced by the
magnetic component of dipole fields. According to this picture, the
disappearance of the EIT-like effect can be explained by the fact
that the magnetic components have the same magnitudes in opposite
directions on the both sides of the upper strip if the structure is
symmetric and thus the induced currents cancel each other out.
Despite their opposite directions, the two magnetic components
cannot be equal in the absence of symmetry so as to produce the
current or quadrupole in the lower pair. This means that the pivot
of the plasmonic EIT is determined by whether the lower pair
(\emph{i.e.}, dark mode) is excited or not.

The point of importance is that the quadrupole can not in general be
excited by normal incidence because of its vanishing dipole moment
(\emph{i.e.}, it is dark). To activate it, a highly angled
illumination must be resorted to.\cite{Zhang2008PRL,MaierNM2009} As
shown in Fig. \ref{fig.3}, two dark modes were magnetically excited
at 120 and 240 THz when the plane wave irradiates on the elongated
pair (790 nm) along the \emph{-y} direction, without symmetry
breaking of both the incident field and the structure itself. The
latter frequency is twice as high as that of the former, which can
be ascribed to be the second-order MPR.\cite{SheridanAPL2007} The
second resonant peak at 240 THz was located coincidentally at the
plasmonic EIT peak as shown in Fig. \ref{fig.2}. Therefore, on the
basis of this physical picture, the lower pair is excited explicitly
at 240 THz.

If we take a closer look at the \emph{z}-component distribution of
the magnetic field at the frequency of plasmonic EIT, as the
physical picture describes, it is unambiguous that the magnetic
fields point in and out of \emph{x-y} plane on both sides of the
upper strip [see Fig. \ref{fig.4}(a)]. Since there is the fact that
the magnetic field is $\pi/2$ out of phase with the electric
field,\cite{Lu2009,BurresiScience2009} the \emph{y}-component
distribution of the electric field is illustrated in Fig.
\ref{fig.4}(b) with a $\pi/2$ phase difference compared to the
magnetic one, where the dark mode is activated and two antisymmetric
quadrupolar resonances are induced on both sides of the upper pair
by two antisymmetric magnetic fields of dipole fields. Duo to the
excitation of the two circular currents, the plasmonic EIT was also
achieved in the symmetric structures. Because the length of the
lower pair was elongated, more room is provided to accommodate the
two circular currents thereby avoiding canceling each other out.
Consequently, the second-order MPR is excited so that the plasmonic
EIT can appear even in the symmetric structures (see Fig.
\ref{fig.2}). This possibility was not even considered as a
candidate for plasmonic EIT.

Furthermore, as shown in Fig. \ref{fig.3}, the intensity of the
second-order MPR is larger than that of the fundamental mode so as
to broaden the FWHM of the plasmonic EIT peak compared with that in
Ref. 11. It is generally believed that the strongly coupling renders
the larger FWHM.\cite{TassinPRL2009} On the other hand, the strength
of the coupling can be tuned by adjusting the vertical distance $h$.
The larger the vertical distance, the smaller the FWHM, of course,
at the cost of transmission.

In conclusion, it is possible for the plasmonic EIT to be realized
even in symmetric structures. This is determined by whether the MPR
can be excited or not, which clarifies the puzzle of the essence of
coupling and strongly validates the mechanism of plasmonic EIT.
Essentially, symmetry or asymmetry is not the pivot of the plasmonic
analogue of EIT, but the MPR is. The resembling modification was not
carried out on the upper strip at the moment because the frequency
scaling law of SPPs is more complicated than that of MPR arising
from the phase shift.\cite{STPRB2008} The point of great interest is
that the optical response is invariant, whereas the restrictions on
size are partly relaxed, and thus the fabrication is much easier.

Yuehui Lu would like to express his gratitude to Nguyen Thanh Tung
for valuable discussions. This work was supported by the MEST/NRF
through the Quantum Photonic Science Research Center, Korea and by
the Research fund of HYU (HYU-2008-T).

\clearpage

\clearpage
\begin{center}
\textbf{Figure captions}
\end{center}

Fig. 1. (a) Three-dimensional view and (b) two-dimensional view of
the unit cell. The geometric parameters are $w = 80$ nm, $d = 100$
nm, $l_{1} = 346$ nm, $l_{2} = 790$ nm, and $s = 0$. The vertical
distance between the upper gold strip and the lower pair of gold
strips is denoted to be $h$ and the thickness of each strip, $t$, is
40 nm. The periods of structure is 870 nm in both the \emph{x} and
\emph{y} directions. The incident plane wave irradiates along the
\emph{z} direction and its electric component, \textbf{E}, is
parallel to the \emph{x} direction.

Fig. 2. (color online) (a) Transmission and (b) absorption spectra
with various vertical distances $h$. The black curves in (a) and (b)
are obtained with the same parameters ($h=70$ nm) except $l_{2} =
315$ nm.

Fig. 3. (a) Schematics for the incident plane wave on the lower pair
of gold strips ($l_2 = 790$ nm), where the wave is parallel to the
strips and its electric field along the \emph{x} direction. The
arrow is an $\emph{E}_{y}$ probe placed 10 nm away from the center
of the end facet. (b) Spectral response of the $\emph{E}_{y}$ probe.

Fig. 4. (a) \emph{z}-component distribution of magnetic field at the
frequency of plasmonic EIT with $h = 70$ nm, where the phase is
$150^{\circ}$. (b) The \emph{y}-component distribution of electric
field at the same frequency, where the phase is $60^{\circ}$.

\clearpage

\clearpage
\begin{figure}[t!]
\includegraphics[width=10.0cm]{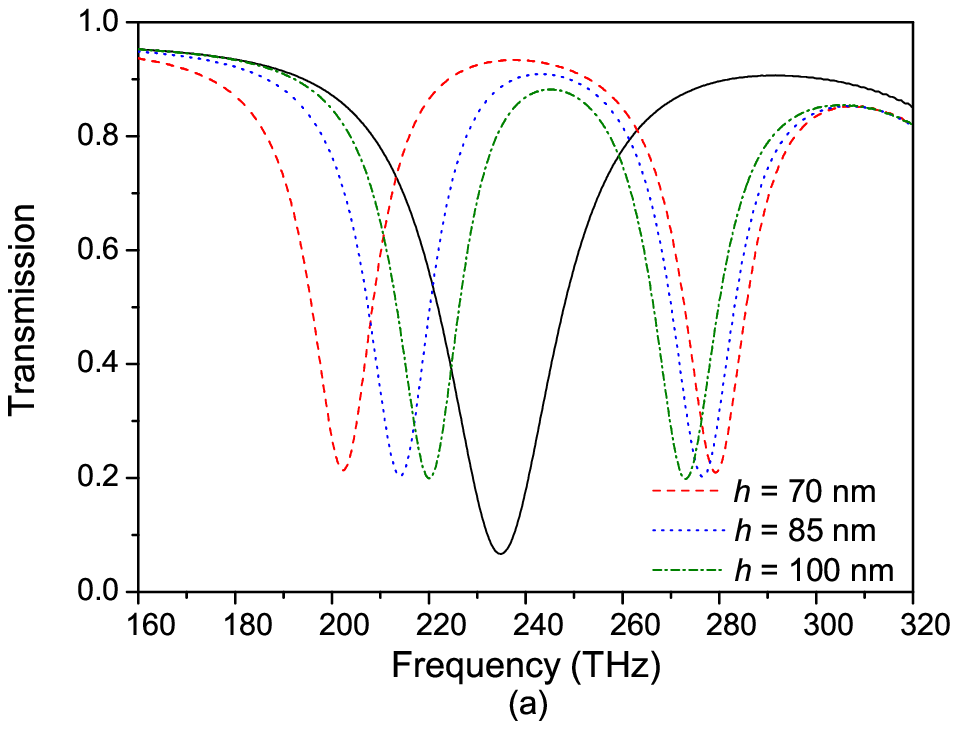}
\end{figure}
\begin{figure}[t!]
\includegraphics[width=10.0cm]{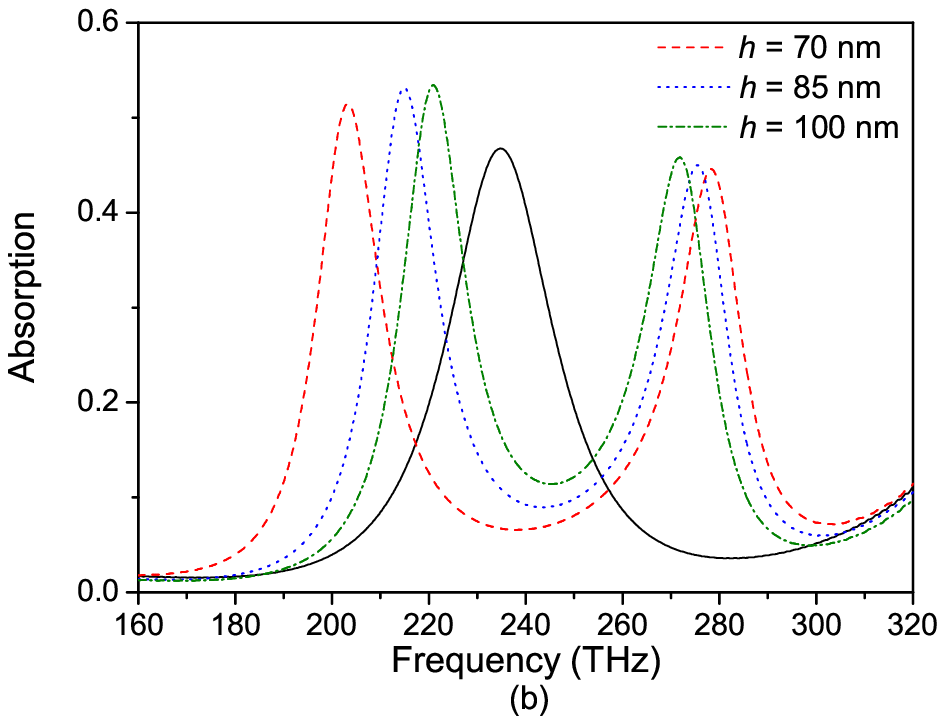}
\caption{Jin \emph{et al.}}\label{fig.2}
\end{figure}

\clearpage
\begin{figure}[t!]
\includegraphics[width=10.0cm]{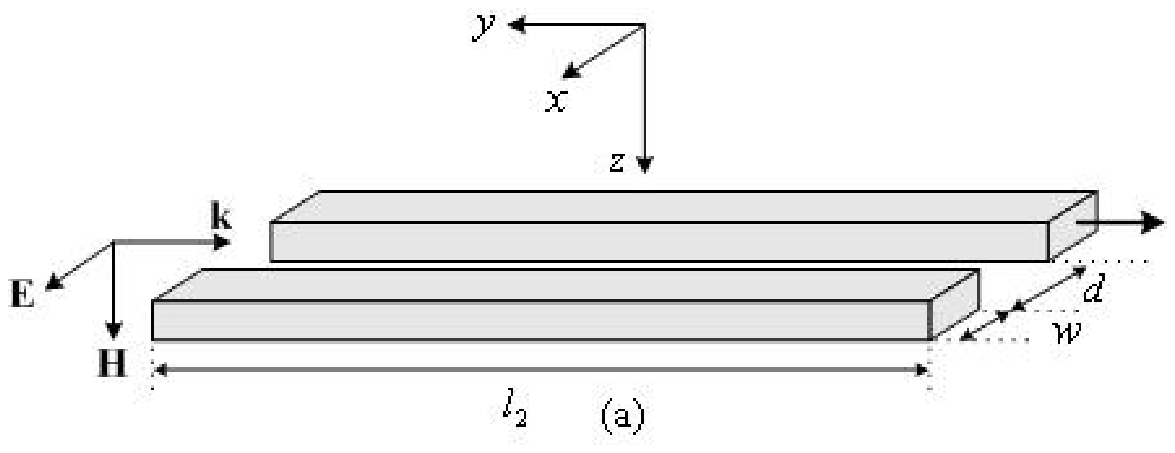}
\end{figure}
\begin{figure}[t!]
\includegraphics[width=10.0cm]{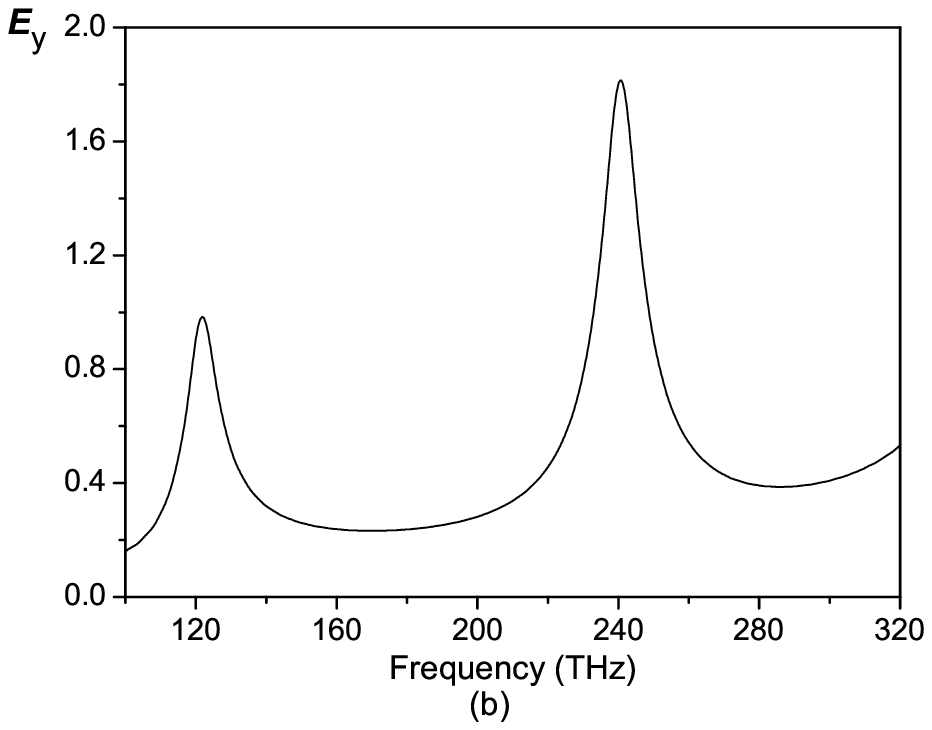}
\caption{Jin \emph{et al.}}\label{fig.3}
\end{figure}

\clearpage
\begin{figure}[t!]
\includegraphics[width=10.0cm]{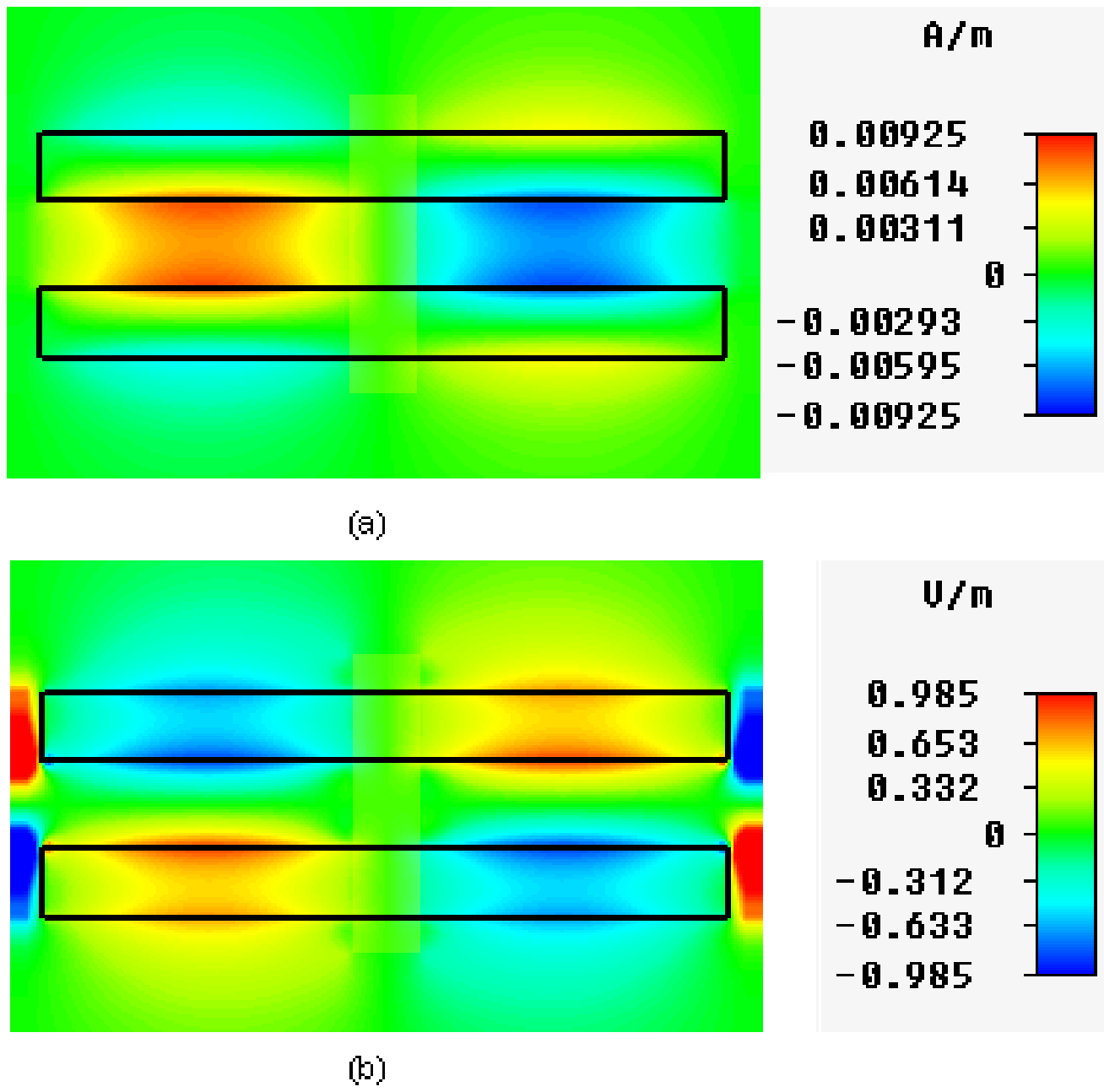}
\caption{Jin \emph{et al.}}\label{fig.4}
\end{figure}

\end{document}